\documentstyle[12pt,epsf]{article}
\newcommand{\be}{\begin{equation}}
\newcommand{\ee}{\end{equation}}

\newcommand{\als}{\mbox{$\alpha_{s}$}}
\newcommand{\s}{\mbox{$\sigma$}}

\newcommand{\bi}[1]{\bibitem{#1}}
\newcommand{\fr}[2]{\frac{#1}{#2}}

\newcommand{\gm}{\mbox{$\gamma_{\mu}$}}

\newcommand{\GD}{\mbox{$\tilde{G}$}}
\newcommand{\gf}{\mbox{$\gamma_{5}$}}

\newcommand{\Ima}{\mbox{Im}}

\begin{document}
\begin{titlepage}
\rightline{\vbox{\halign{&#\hfil\cr
&UQAM-PHE-97/02\cr
&\today\cr}}}
\vspace{0.5in}
\begin{center}

\large{\bf  
Constrained MSSM and the electric dipole moment of the neutron}
\\
\medskip
\vskip0.5in

\normalsize {{\bf C. Hamzaoui}$^{\rm a}$\footnote{hamzaoui@mercure.phy.uqam.ca}, 
{\bf M. Pospelov}$^{\rm a,b}$}\footnote{pospelov@mercure.phy.uqam.ca} 
and 
{\bf R. Roiban}$^{\rm a}$\footnote{radu@mercure.phy.uqam.ca}
\smallskip
\medskip

{ \sl $^a$
D\'{e}partement de Physique, Universit\'{e} du Qu\'{e}bec a 
Montr\'{e}al\\ 
C.P. 8888, Succ. Centre-Ville, Montr\'{e}al, Qu\'{e}bec, 
Canada, H3C 3P8\\
$^b$Budker Institute of Nuclear Physics, Novosibirsk, 
630090, Russia} 
\smallskip
\end{center}
\vskip1.0in

\noindent{\large\bf Abstract}
\smallskip\newline
We study
the constraints on the CP-violating 
soft-breaking phases in the 
minimal supersymmetric standard model using the 
limits on the chromoelectric 
dipole moment of the strange quark extracted 
from the neutron EDM experiment. 
Our investigation shows that 
the phase mediated by the gluino exchange 
diagram  has to be very small, 
$\phi\leq 8\cdot 10^{-4}$, for the common 
supersymmetric mass of the order of 100 GeV. 
Then, solving the renormalization group equations 
analytically by iterations, we calculate 
the  electric dipole moment of the neutron 
in the MSSM with CP-conserving 
soft-breaking parameters for the case of 
three and four generations. For the
three-generation case we resolve the apparent 
discrepancies between 
order-of-magnitude estimates and numerical 
calculations existing in the literature. 
In this case the EDM of the neutron does not exceed 
$10^{-32} e\cdot cm$. For the 
four-generation case we show that there is a 
significant enhancement which renders the EDM 
of the neutron at a measurable level 
of $10^{-26}e\cdot cm$. 

\end{titlepage}

\baselineskip=20pt

\newpage
\pagenumbering{arabic}
\section{\bf Introduction}
\label{intro}

The current experimental
limit on the electric dipole moment (EDM) of the neutron 
\cite{LG}, \hfil\break
$d_N/e<10^{-25}\,cm$, exceeds the realistic Standard Model
prediction for this quantity by seven orders of magnitude.
This gap between theory and experiment provides an excellent 
opportunity to 
limit a new CP-violating physics beyond the Standard 
Model (SM). The purpose of 
this work is to consider the electric dipole moment 
of the neutron in the 
Minimal Supersymmetric Standard Model (MSSM) with 
three and four generations of 
quarks. 

The Minimal Supersymmetric Standard Model~\cite{habnil} 
looks nowadays as a very promising 
candidate for the physics 
beyond the SM. It can be
probed experimentally through the  high energy 
production of superpartners. 
Since no supersymmetric particles have been found 
so far, one can limit the parameter 
space of MSSM using the supersymmetric 
radiative corrections to the low-energy 
observables. The limits from the precise 
measurements of EDMs of the neutron, 
electron and heavy atoms are known to be 
a very good source of information in this respect 
\cite{MSSM-EDM}. Namely, the one-loop 
contribution to the electric dipole 
moments of quarks allows one to put stringent 
limits on the CP-violating phases 
in the soft-breaking sector, 
$\phi\leq \,{\rm few}\times 10^{-3}$\cite{MSSM-EDM}. 
Alternatively, 
the squark masses have to lie in the TeV range if 
we believe that the phases are 
not suppressed. This dilemma is often referred to 
as the Supersymmetric CP problem. 

To avoid this problem from the very beginning, one 
can postulate that the 
soft-breaking sector as well as the $\mu$-term in 
the superpotential are 
CP-conserving. This choice of parameters is often 
suggested by supergravity 
\cite{DGH}. In addition to that, it is reasonable 
to assume the universality in 
the mixing of left- and right-handed scalar quarks 
to avoid unwanted FCNC at low 
energies. As a result, the only source of 
CP-violation resides in the Yukawa 
couplings and can be described at the electroweak 
scale by the Kobayashi-Maskawa 
(KM) phase. In principle, the prediction for the neutron EDM 
in MSSM can be different from the SM 
one due to the specific supesymmetric 
contributions and the 
KM-related phase which appears in the squark mass 
matrix \cite{Dunc}. 

The order-of-magnitude estimates of EDMs \cite{DGH,RS}, 
handling the main 
dependence on Yukawa couplings and mixing angles, 
suggest that the EDM of the 
neutron does not exceed $10^{-32} e\cdot cm$. This 
means that in this type of 
models the contribution from supersymmetric loops 
never exceeds the 
long-distance contribution to EDM from the usual 
SM \cite{KhZ}. To improve the 
order-of-magnitude estimates, one has to solve the 
nonlinear renormalization group 
equations which is difficult to do analytically. 
Surprizingly enough, the 
attempts to solve these equations numerically by plugging 
in the values for quark 
masses and mixing angles measured at the electroweak 
scale give somewhat bigger 
results \cite{Ital,Japan,Engl}. The discrepancy between 
\cite{RS} and \cite{Japan}, 
for example, constitutes six orders of 
magnitude so that it 
definitely requires an explanation. 
To resolve the apparent discrepancies and
find the connection between the two approaches, 
we solve the renormalization group
equations analytically, assuming that the renormalization 
group coefficient, $t=(4\pi)^{-2}\log(\Lambda^2/M_W^2)$, is small.
In this case the equations are solvable by iterations and an 
analytical result for EDM can be obtained.
It turns out that a
finite result arises already after the first 
iteration, so that $d_N$ is 
proportional  to the first power of $t$. This 
result may be viewed as the 
intermediate step between parametrical estimates and 
numerical calculations. The 
fact that $t$ is not small can change the final answer 
somehow, but not by many 
orders of magnitude if we believe that all Yukawa 
couplings remain in the 
perturbative regime in the whole interval from 
$M_W$ to $\Lambda$. In any case, 
this question is more of methodological 
interest because of the expected 
smallness of the result. 

As it was shown in \cite{HP,HP1}, the existence of 
a possible fourth 
generation in the framework of the usual SM gives 
a significant enhancement to 
EDM. The result, however, is still too small to be 
probed experimentally. With the 
natural assumptions about mixing angles it does 
not exceed $10^{-29} e\cdot cm$ 
\cite{HP1}. In this paper we consider the
MSSM with four generations 
and show that the value of the EDMs is 
significantly enhanced. This may 
give nontrivial constraints on 
the parameters of the model, complementary to those 
coming from the direct 
search of particles from the fourth generation and 
from the analysis of the 
renormalization group evolution of Yukawa couplings \cite{GMP,CHW}. 

We organize the paper as follows.
In the next section we obtain 
new, apparently stronger, limits on CP-violating 
phases in the soft breaking 
sector mediated by the chromoelectric dipole 
moment (CEDM) of the strange quark 
and present the analytical calculation of EDM in 
the constrained MSSM with three 
generations. In Section 3 we estimate 
the EDMs in the MSSM with four generations. 
Our conclusions are sumarized in Section 4.

\section{EDM in MSSM with three generations}

We start by writing down the structure of 
the CP-odd effective Lagrangian dim$\le 6$ at 
the scale of $1$ GeV which gives a 
major contribution to the EDM of the neutron:
\be
{\cal L}_{eff}(x)=\theta\fr{\als}{8\pi}
G^a_{\mu\nu}\GD^a_{\mu\nu}+
i\fr{c_W}{6}g_s^3f^{abc}\GD^a_{\alpha\beta}
G^b_{\beta\mu}G^c_{\mu\alpha}+
i\sum_i\fr{\tilde{d}_i}{2}\bar{q}_it^a(G^a\s)\gf q_i+
i\sum_i\fr{d_i}{2}\bar{q}_i(F\s)\gf q_i,
\label{eq:Leff}
\ee
where the sum over $i$ runs over light flavours, 
$u$, $d$ and $s$. All other 
possible CP-odd operators dim=6 are irrelevant 
(see, for example, Ref. 
\cite{HP}). In SM we have to take into account 
also the so called long-distance 
contributions: the combination of two 
flavour-changing operators saturated by all 
possible hadronic states \cite{KhZ}. 

Here we concentrate ourselves on the calculation of 
EDMs, $d_i$, and colour EDMs 
$\tilde{d}_i$. The relative meaning of these operators 
for the electric dipole 
moment of the neutron is quite a controversial subject.
To get the final result, one has to know how to
calculate the EDM of the neutron in
terms of $d_i$ and $\tilde d_i$. This is a very
difficult task if one takes into account the
strong dynamics at large distances.
To simplify this task, the naive nonrelativistic 
quark model is used very often. It 
relates $d_N$ with $d_u$ and $d_d$ in a simple manner:
\be
d_N=\fr{4}{3}d_d-\fr{1}{3}d_u
\label{eq:naiv}
\ee
At the same time, the chromoelectric dipole moments 
as well as the electric 
dipole moment of s-quark are believed to be 
suppressed. The importance of the
information provided by the EDM data suggests, 
however, that the naive formula 
based on the non-relativistic quark model must 
be abandoned and replaced with more elaborated 
approaches handling long-distance dynamics.

Let us suppose for a moment that we have the 
approximate universality in the 
sector of EDMs and CEDMs, for example, 
$d_d/(em_d)\simeq d_s/(em_s)\simeq a$, 
$\tilde d_d/(m_d)\simeq \tilde d_s/(m_s)\simeq b$ 
and $a\sim b/g_s$. In a 
recent work \cite{EF}, the contribution of different 
quark EDMs to the EDM of the
neutron were treated using the proton spin 
experimental data implying a 
non-zero content of the strange quark spin in 
the nucleon. In other words, the 
authors of Ref. \cite{EF} assume the identity 
between axial, $(\Delta q)_A$, and 
tensor, $(\Delta q)_T$, charges. (The definitions 
for these quantities are: $\bar 
N \gm\gf N(\Delta q)_A=\langle N|\bar q \gm\gf q|N\rangle$;  
$\bar 
N\sigma_{\mu\nu}N(\Delta q)_T=\langle N|\bar q \sigma_{\mu\nu} 
q|N\rangle$\footnote{ The EDM Lorentz structure is 
reducible to $O_T$ using the 
identity $2\sigma_{\mu\nu}\gf= 
i\epsilon_{\mu\nu\alpha\beta}\sigma_{\alpha\beta}$}.) 
As a result, a sort of 
cancellation between strange and down quark 
contributions to neutron EDM was 
observed. This leads to apparently milder 
limits on $\phi$ \cite{EF}. It is 
clear, however, that the axial and tensor
charges correspond to different structure 
functions and do not have to 
coincide \cite{JJ}. It is especially true 
for the tensor charge of the s-quark 
in the neutron. Representing basically the 
sea quark contribution, the matrix 
element of the strange quark should be especially 
sensitive to the property of 
the operator with respect to charge conjugation. 
Since $C(\bar q \gm\gf 
q)=+\bar q \gm\gf q$ and $C(\bar q 
\sigma_{\mu\nu} q)=-\bar q \sigma_{\mu\nu} 
q$, we deduce that $(\Delta s)_T<(\Delta s)_A$. 
The calculation in the 
instanton-inspired model shows, in fact, that 
$(\Delta s)_T\ll(\Delta s)_A$ 
\cite{Pol}. Taking into account the results of 
the tensor charge calculation on 
the lattice \cite{Hats}, within QCD sum rules 
\cite{Ji} and in 
instanton-inspired model \cite{Pol}, we 
conclude that in the quark EDM channel 
the naive formula (\ref{eq:naiv}) is roughly 
correct and, at the same time, the 
electric dipole moment of the s-quark is 
unimportant providing {\em no} 
conspiracy among different flavours.  

In our opinion, the question of conspiracy among 
the EDMs of different flavours 
is not relevant in MSSM on account of large 
contributions coming from the 
chromoelectric dipole moments of quarks. We 
would like to apply here the results 
obtained in Refs. \cite{KKZ,KKY,KK,KZy} in the context 
of chiral perturbation 
theory and QCD sum rules. If the above mentioned 
universality is held, the 
contributions from the chromoelectric dipole 
moments to the neutron EDM tend to 
dominate over the quark EDM contributions. The 
resulting estimate is given by \cite{KKZ,KK,KZy}:
\be
d_N\simeq e(0.7\tilde d_d + 0.1 \tilde d_s)
\label{eq:ds}
\ee
and normally $e\tilde d_i>d_i$ due to the large $g_s$ 
factor in $\tilde d_i$. 
Moreover, the second term in the estimate (\ref{eq:ds}) 
tends to dominate due to the large $m_s$. 
Simple arguments favoring the suppression of the strange 
quark contribution in the quark EDM channel do not work 
for CEDM contributions. 
(For a recent discussion on the strange quark matrix 
elements in the nucleon 
see, for example, Ref. \cite{Z}.)

It was shown recently that this operator produces a 
large neutron EDM in the 
supersymmetric $SO(10)$ model \cite{KZy}. In the 
Minimal Supersymmetric Standard 
Model with CP-violating phases in the soft-breaking 
sector, the gluino 
exchange diagram leads to the following CEDM of a 
quark:
\be
\tilde d_i\simeq g_s\fr{5\alpha_s}{72\pi}~
\fr{m_i}{m^3}~F(m_{\lambda_3}^2/m^2)~(A\sin\phi_A
+\mu\tan\beta\sin\phi_B),
\label{eq:CEDM}
\ee
where $\phi_A=\Ima (Am^*_{\lambda_3})$ and $\phi_B=
\Ima (\mu m^*_{\lambda_3})$ are the specific 
CP-violating phases which 
appear in this diagram. The invariant function in 
Eq. (\ref{eq:CEDM}) contains the dependence of the 
supersymmetric masses in the loop. It is normalized 
in such a way that $F(1)=1$. Eq. (\ref{eq:CEDM})
gives us the limits on the CP-violating 
phases considerably stronger than those coming from 
the EDMs of quarks. 
Taking $m_s(100\,{\em GeV})=80\,{\em MeV}$, 
$\mu\sim A\sim m$ and 
using (\ref{eq:ds}), (\ref{eq:CEDM}) and experimental 
limits we 
obtain the following constraint on the CP-violating phase:
\be
\phi_{A(B)}\left(\fr{100\mbox{GeV}}{m}\right)^2\leq 
8\cdot 10^{-4}.
\label{eq:phi}
\ee
When obtaining the limit (\ref{eq:phi}) we 
have assumed that both phases can be constrained 
independently, i.e. no fine tunning among $A$, 
$\mu\tan\beta$, $\phi_A$ and $\phi_B$. It is interesting 
to note also that the function $F$ reaches its maximum, 
$F\sim 2.7$ for $m_{\lambda_3}\simeq 0.2\,m$, so that 
the constraints on the phases in this case are almost 
three times stronger.

These limits are very restrictive and from naturalness 
reasons we assume further on that the soft 
breaking sector does respect CP-symmetry. 
In what follows, we calculate~SUSY contributions 
to the coefficients in (\ref{eq:Leff}) 
due to the presence of CP-violation in Yukawa couplings.

We take the superpotential of MSSM in the following standard form:
\be
W=\bar{U}{\bf Y}_uQH_u+\bar{D}{\bf Y}_dQH_d+
\bar{E}{\bf Y}_eLH_d+\mu H_uH_d
\ee

The soft breaking sector at the unification scale comprises flavour-blind 
scalar mass terms:
\be
m_{H_d}^2|H_d|^2+m_{H_u}^2|H_u|^2 + m^2_{U}\tilde{U}^{\dagger}\tilde U
+m^2_{D}\tilde{D}^{\dagger}\tilde D+m^2_{Q}\tilde{Q}^{\dagger}\tilde Q~+H.c.;
\ee
so called $A$-terms proportional to Yukawa couplings
\be
A\left(\tilde{U}^{\dagger}_R{\bf Y}_u\tilde{Q}_LH_u+
\tilde{D}^{\dagger}_R{\bf Y}_u\tilde{Q}_LH_d+
\tilde{D}^{\dagger}_R{\bf Y}_u\tilde{Q}_LH_d\right)~+H.c.;
\label{eq:A}
\ee
B term,
\be
B\mu H_u\!\cdot\! H_d+H.c.;
\ee
and the gaugino masses
\be
\sum_{i=1,2,3}m_{\lambda_i}
\bar{\lambda}_{iR}\lambda_{iL}~+H.c.
\ee
In formal language, the absence of additional 
CP-violating phases  means the following:
\be
A=A^*,~B=B^*,~\mu=\mu^*,~m_{\lambda_i}=m_{\lambda_i}^*.
\ee

The squark mass matrices at the unification scale
\begin{eqnarray}
M_{\tilde u}^2 = \left(\begin{array}{cc}
m_{Q}^2+{\bf M}_u^{\dagger}{\bf M}_u&
(A+\mu\cot\beta) {\bf M}_u^{\dagger}\\
\noalign{\vskip10pt}
{\bf M}_u(A+\mu\cot\beta) &
m^2_{U}+{\bf M}_u{\bf M}_u^{\dagger}
\end{array} \right)
\label{eq:Mu}
\end{eqnarray}
\begin{eqnarray}
M_{\tilde d}^2 = \left(\begin{array}{cc}
m_{Q}^2+{\bf M}_d^{\dagger}{\bf M}_d&
(A+\mu\tan\beta) {\bf M}_d^{\dagger}\\
\noalign{\vskip10pt}
{\bf M}_d(A+\mu\tan\beta)&
m^2_{D}+{\bf M}_d{\bf M}_d^{\dagger}
\end{array} \right)
\label{eq:Md}
\end{eqnarray}
do not lead to any flavour-changing effects, not 
mentioning CP-violation, since 
they are diagonalizable in the generation space 
by the same bi-unitary 
transformation as quark mass matrices. As a result, 
quark-squark-neutralino or 
quark-squark-chargino interactions do not develop 
flavour-changing vertices. The terms
bilinear in ${\bf M}_u$ and ${\bf M}_d$ in the squark mass 
matrices (\ref{eq:Mu}) and 
(\ref{eq:Md}) originate from the $F$-term in the 
superpotential. Since they are 
not related to the renormalization group running from large 
scale down to $M_W$, we shall 
refer to them as the tree-level terms. The 
renormalization group 
evaluation of the soft-breaking parameters 
and Yukawa couplings down to the electroweak 
scale indroduces 
flavour-changing entries and 
brings, for example, the dependence of ${\bf Y}_d$ 
in $M_{\tilde u}^2$ 
\cite{Dunc}. At the scale of the electroweak 
symmetry breaking the squark mass 
matrices can be parametrized as follows:
\begin{eqnarray}
M_{\tilde u}^2 = \left(\begin{array}{cc}
M^2_{uLL}&
M^2_{uLR}\\
\noalign{\vskip10pt}
M^2_{uRL} &
M^2_{uRR}
\end{array} \right)~
M_{\tilde d}^2 = \left(\begin{array}{cc}
M^2_{dLL}&
M^2_{dLR}\\
\noalign{\vskip10pt}
M^2_{dRL} &
M^2_{dRR}
\end{array} \right)
\label{eq:M1}
\end{eqnarray}
where different blocks have the following meaning:
\begin{eqnarray}
\nonumber
M^2_{uLL} &=& m^2_{Q1}+{\bf M}_u^{\dagger}{\bf M}_u
+c_1{\bf Y}_u^{\dagger}{\bf Y}_u\nonumber+f_1{\bf Y}_d^{\dagger}{\bf Y}_d
+c_2({\bf Y}_u^{\dagger}{\bf Y}_u)^2+f_2({\bf Y}_d^{\dagger}{\bf Y}_d)^2+\nonumber\\
& &+c_2' \{{\bf Y}_u^{\dagger}{\bf Y}_u,{\bf Y}_d^{\dagger}{\bf Y}_d\}+...\nonumber\\
M^2_{uLR} &=& (M^2_{uRL})^\dagger=(A_1+\mu\cot\beta) 
{\bf M}_u^{\dagger}+A(h_1{\bf Y}_d^\dagger {\bf Y}_d+k_1{\bf Y}_u^{\dagger}{\bf Y}_u+...)
{\bf M}_u^{\dagger}\nonumber\\
M^2_{uRR} &=&m^2_{u1}+{\bf M}_u{\bf M}_u^{\dagger}+
l_1{\bf Y}_u{\bf Y}_u^{\dagger}+l_2({\bf Y}_u{\bf Y}_u^{\dagger})^2+
l_2'{\bf Y}_u{\bf Y}_d^{\dagger}{\bf Y}_d{\bf Y}_u^{\dagger}+...
\label{eq:I1}
\end{eqnarray}
and similarly for the blocks of $M_{\tilde d}^2$. 
The explicit form of all coefficients can
be found in the Appendix. In the expression 
(\ref{eq:I1}) all quark masses and Yukawa couplings 
are taken at the electroweak scale. 
The dots stand for other terms in the series of the 
increasing power of Yukawa 
couplings and renormalization group factor $t$.  The subscripts 
of the coefficients 
$c_1,~c_2,~f_1,...$ denote the smallest power of $t$ 
in which these entries to 
the mass matrix can arise, i.e. $f_1\sim t^1$, 
$f_2\sim t^2$, etc. Please, note 
also that $c_2$, $f_2$, etc. {\em are not} the 
contributions from the two-loop 
beta function. Finally, $A_1$, $m^2_{Q1}$ and 
$m^2_{U1}$ are soft-breaking 
parameters renormalized by gauge interactions and 
by Yukawa interactions 
effectively conserving flavour, i.e. proportional 
to Tr$\big[{\bf Y}_u{\bf Y}_u^{\dagger}\big]$.

As a result, one-loop diagrams with chargino and gluino 
inside the loop, Fig.~1 and Fig.~2, can develop 
imaginary parts and thus lead to a nonvanishing EDM. 
This effect, however, is severely suppressed by the 
combination of Yukawa couplings and mixing angles. 
The extraction of the 
imaginary part gives almost the structure of the 
Jarlskog invariant 
\cite{Jarlsk} and leads to the following simple 
estimate \cite{DGH}:
\begin{eqnarray}
d_d\sim e(\mbox{loop factors})\times
\fr{1}{m^2}\Ima\bigl[{\bf M}_d
{\bf Y}_u^4{\bf Y}_d^2{\bf Y}_u^2\bigl]_{11}
\sim e (\mbox{loop f.})\times J_{CP}\fr{m_d}{m^2}
Y_t^4Y_c^2Y_b^2.
\label{eq:Jarl}
\end{eqnarray}
where $J_{CP}=\Ima(V^*_{td}V^{}_{tb}V^*_{cb}V^{}_{cd})$; 
$m^2$ is the characteristic 
momentum in the loop of the order of supersymmetric 
masses; "loop factors" 
denotes numerical coefficients reflecting the loop 
origin of the effect and the subscript $11$ denotes the 
projection on the initial flavour $d$. Even 
with optimistic expectations for the numerical 
coefficients in Eq. (\ref{eq:Jarl}), the result can 
hardly exceed $10^{-34} e\cdot cm$ for 
$\tan\beta\sim 1$. At the same time pure gluonic 
operators, as well as the EDM 
of the u-quark, are further 
suppressed by Yukawa couplings coming from down 
quark sector. 

To resolve apparent discrepancies with the numerical 
calculations, we solve the 
renormalization group equation analytically by 
iterations retaining the smallest 
power of $t$. This procedure is very simple; it 
brings analytical answers for all 
operators in Eq. (\ref{eq:Leff}) and at the same 
time it gives the possibility 
to check all numerical calculation in an easy way. 

In both papers quoting the estimate (\ref{eq:Jarl}), 
Ref.\cite{DGH} and 
Ref.\cite{RS}, the effect is claimed to arize in the 
third or fourth order in 
renormalization group coefficient $t$. Surprizingly enough, we 
arrived to somewhat 
different conclusion. The chargino exchange diagram 
gives a nonvanishing $d_d$ 
in the lowest possible (first) order in $t$. Clearly, 
this contribution is related 
with the ${\bf M}_u$-dependence of the u-squark mass 
matrix (\ref{eq:M1}) coming from 
the tree-level and is not associated with $t$. The 
gluino exchange diagram yields 
EDMs in the down-quark sector in the third order in $t$. 

\medskip
\medskip
{\bf Chargino contribution}\newline
The chargino exchange diagram which contributes 
to EDMs and CEDMs in the 
down-quark sector contains the u-squark line. 
The result of the renormalization group running, up to 
the first power in $t$ is 
summarized by the set of formulae displayed in 
the Appendix. Here we give the 
truncated form and list only relevant entries 
in (\ref{eq:I1}) with 
$Y_d^2$-dependence:
\begin{eqnarray}
M_{\tilde u}^2 &= &\left(\begin{array}{cc}
m_{Q1}^2+{\bf M}_u^{\dagger}{\bf M}_u+
f_1{\bf Y}_d^\dagger {\bf Y}_d&
(A_1+\mu\cot\beta) {\bf M}_u^{\dagger}+
Ah_1{\bf Y}_d^\dagger {\bf Y}_d{\bf M}_u^{\dagger}\\
\noalign{\vskip10pt}
(A_1+\mu\cot\beta){\bf M}_u +
Ah_1{\bf M}_u{\bf Y}_d^\dagger {\bf Y}_d &
m^2_{U1}+{\bf M}_u{\bf M}_u^{\dagger}
\end{array} \right)\\
\nonumber
\end{eqnarray}
where
\begin{eqnarray}
f_1&=&-\frac{\log(\Lambda^2/m^2)}{16\pi^2}
\Bigl[m_{Q}^2+m_{U}^2+m_{H_u}^2 +A^2\Bigl];\;\;\;\;
h_1=-\fr{\log(\Lambda^2/m^2)}{16\pi^2}
\label{eq:Mu1}
\end{eqnarray}
Assuming the equality between $m_{U}$ and $m_{Q}$ 
at the high-energy scale, 
$m_{U}=m_{Q}=m$, we can conclude that the difference 
of $m_{U1}$ and $m_{Q1}$ is 
also proportional to $t$. 

The next step could be the expansion of the squark 
propagator in $A{\bf M}_u$ and ${\bf M}_u^2$ 
and the extraction of the CP-violating part from the 
string of Yukawa couplings and 
mass matrices \cite{DGH}. This procedure is completely 
justified for the charm 
quark mass since $m_c^2,~Am_c\ll m^2$ and seems to be 
ill-defined for the top 
flavour since we expect $m_t^2,~Am_t\sim m^2$. Fortunately, there is 
no need to expand the squark propagator in terms of $m_t^2$. 
Let us take the first 
term of expansion in 
$A$. Then the resulting flavour structure of the squark 
line has the following simple form:
\begin{eqnarray}
\Ima \left[ \fr{1}{p^2-m_{Q1}^2-{\bf M}_u^{\dagger}{\bf M}_u}
f_1{\bf Y}_d^\dagger {\bf Y}_d
\fr{A_1+\mu\cot\beta}{p^2-m_{Q1}^2-{\bf M}_u^{\dagger}{\bf M}_u}
{\bf M}_u^\dagger
\fr{1}{p^2-m_{U1}^2-
{\bf M}_u{\bf M}_u^{\dagger}}{\bf M}_u \right.+\nonumber\\
\left. \fr{1}{p^2-m_{Q1}^2-{\bf M}_u^{\dagger}{\bf M}_u}
Ah_1{\bf Y}_d^\dagger {\bf Y}_d{\bf M}_u^{\dagger}
\fr{1}{p^2-m_{U1}^2-
{\bf M}_u{\bf M}_u^{\dagger}}{\bf M}_u\right]_{11}
\simeq\nonumber\\
\simeq -J_{CP}m_c^2 m_t^4 Y_b^2
\fr{f_1(A+\mu\cot\beta)-
A h_1(m_{Q1}^2-m_{U1}^2)}{(p^2-m^2)^3(p^2-m^2-m_t^2)^2}
\label{eq:fs}
\end{eqnarray}
where we took the difference of the two soft-breaking 
masses to be much smaller 
than the masses themselves and neglected $m_u^2$ and 
$m_c^2$ in the denominator. It is easy to see 
that the term 
proportional to $h_1$ cannot develop any CP-violating part 
unless $m_{U1}\neq m_{Q1}$. However, since we are interested
only in the first order in $t$, this contribution can
be neglected being of the order $t^2$. 
In order to get the exact dependence of $m_t$, one has to sum 
up the series in $(A+\mu\cot\beta)m_t$, i.e. to 
take into account large $\tilde t_L-\tilde t_R$ mixing. 

This complication as well as the effect of 
wino-higgsino mixing are irrelevant 
for the question under study and we would like 
to neglect them. Assuming 
also for simplicity that char- gino mass is 
equal to the scalar quark mass $m$ and taking 
$f_1\simeq -3m^2(16\pi^2)^{-1}\log(\Lambda^2/m^2)$, 
after the trivial integration, we arrive at 
the following simple result:
\begin{eqnarray}
d_d&\simeq&\fr{1}{5}\fr{1}{16\pi^2}
(\fr{5}{7}e_d-e_{\tilde\chi})
\frac{\log(\Lambda^2/m^2)}{16\pi^2}\,J_{CP}
\frac{m_d m_c^2m_t^4Y_b^2}{m^7}\,
\frac{A+\mu\cot\beta}{v_uv_d}~~\mbox{for}~m\gg m_t\nonumber
\\d_d&\simeq&\fr{1}{2}\fr{1}{16\pi^2}
(\fr{3}{5}e_d-e_{\tilde\chi})
\frac{\log(\Lambda^2/m^2)}{16\pi^2}\,J_{CP}
\frac{m_d m_c^2Y_b^2}{m^3}\,
\frac{A+\mu\cot\beta}{v_uv_d}~~~~~~\hskip-.75truept
\mbox{for}~m\ll m_t
\label{eq:res}
\end{eqnarray}
Here we keep explicit the loop origin of $Y_b^2$ and 
the tree-level origin of 
$m_c^2m_t^4$. In a more general case with 
different supersymmetric 
masses, chargino mixing and realistic 
$m_t$-dependence taken into account, the 
numerical coefficients and the dependence 
of $m_t^2$ and $m^2$ should be 
substituted by some more complicated invariant 
function. Its exact form is beyond 
the scope of our interest. For the 
chromoelectric dipole moment we have a 
similar result:
\begin{eqnarray}
\tilde d_d&\simeq&\fr{g_3}{7}\fr{1}{16\pi^2}
\frac{\log(\Lambda^2/m^2)}{16\pi^2}\,J_{CP}
\frac{m_d m_c^2m_t^4Y_b^2}{m^7}\,
\frac{A+\mu\cot\beta}{v_uv_d}~~\mbox{for}~m\gg m_t\nonumber\\
\tilde d_d&\simeq&\fr{3g_3}{10}\fr{1}{16\pi^2}
\frac{\log(\Lambda^2/m^2)}{16\pi^2}
\,J_{CP}
\frac{m_d m_c^2Y_b^2}{m^3}\,
\frac{A+\mu\cot\beta}{v_uv_d}~~~~\hskip1.2truept
\mbox{for}~m\ll m_t
\end{eqnarray}
Numerically, the EDM of d-quark (\ref{eq:res}) constitutes
\be
|d_d|\sim 6\cdot 10^{-34} 
\left(\fr{\tan\beta}{10}\right)^3e\cdot cm
\label{eq:nedm}
\ee
were we took $J_{CP}\!\simeq\! 2\times 10^{-5}$, 
$\tan\beta>1$, 
$(16\pi^2)^{-1}\log(\Lambda^2/m^2)\!\sim\! 1/3$ 
and $m\!\sim\! A\!\sim\! 100$~GeV. It 
should be noted here that all Yukawa couplings and  
quark masses are taken at 
the scale of $100$ GeV so that the result is 
further suppressed by the QCD 
evaluation of $m_c^2m_b^2$ from $1$~GeV to 
this scale. Even taking into account 
the enhancement associated with the CEDM of 
the strange quark we cannot obtain 
the EDM of the neutron bigger than $10^{-32}e\cdot cm$. 

The numerical enhancement of $d_d$, up to the 
level of $10^{-28} e\cdot cm$,  
observed in Ref. \cite{Japan} is nothing but 
an artifact resulting from a quite 
surprizing choice of the parameters 
$m_{scalar}=100$ GeV and $A=5$ TeV, so that 
$A/m_{scalar}=50$. (This was noted also in Ref. \cite{Engl}.) 
We believe, however, that this choice of parameters 
is highly unnatural if not strictly 
forbidden. Here we would like to show how it 
could change the result 
numerically. In this case, the expansion in 
$Am_c/m_{scalar}^2$ is not 
associated with any numerical smallness since 
this factor is now roughly of the 
order 1. In consequence, the result would not be suppressed 
in $m_c^2$ at all, leading to an
enhancement of the order $10^4$.  

\medskip
\medskip

{\bf Gluino contribution}\newline
It is clear that the result can arise here only 
in the third order in $t$. The 
simple solution for the down squark mass matrix, 
analogous to (\ref{eq:I1}) is 
not sufficient, though. Indeed, it can be shown
 that the flavour-diagonal 
projection of the combination 
\be
\fr{1}{p^2-m_{Q1}^2-(\fr{v_d^2}{2}+c_1)
{\bf Y}_d^\dagger {\bf Y}_d-f_1{\bf Y}_u^\dagger 
{\bf Y}_u}(A'+\mu\cot\beta+
h_1{\bf Y}_u^\dagger {\bf Y}_u){\bf M}_d^\dagger
\fr{1}{p^2-m_{U1}^2-(\fr{v_d^2}{2}+l_1)
{\bf Y}_d{\bf Y}_d^\dagger}
\ee
does not develop any CP-violating part at all. 
This shows that the simple 
estimate presented in Refs. \cite{DGH} and 
\cite{RS} based on the quadratic 
anzatz of the mass matrices are incorrect 
since the exact result in this 
approximation is just zero in any order in $t$. 
This situation resembles the 
cancellation of the two-loop contribution to the 
EDM of a quark in the SM. The combination of different 
diagrams, nonzero by themselves, vanishes after the 
complete summation over all flavours \cite{Shab}. 

To be rigorous, however, we have to go 
further and calculate quartic 
combinations of Yukawa couplings in the squark 
masses and $A$-parameter 
proportional to the second power of $t$, i.e the 
coefficients $c_2$, $d_2$, etc. 
It can be shown also that the $t^3$-order entries 
in the quark mass matrices 
contribute to EDMs in the next (fourth) order in 
$t$ and thus are irrelevant. The 
calculation is very simple and all relevant 
coefficients are listed in the 
Apendix. Omitting all intermediate steps, we 
would like to quote here the final result:
\begin{equation}
d_d\,=\,e_d\,m_d\,m_b^2Y_c^2Y_t^4\,J_{CP}\,
\frac{2\alpha_s}{3\pi}
\,\,\frac{t^3\,A^2}
{m^7}\,\Bigg[\frac{4}{30}A-
(A+\mu\,{\rm tan}\beta)\frac{8A(A+\mu\,{\rm tan}\beta)-
(3m^2+A^2)}
{42m^2}\Bigg]
\end{equation}
It is obtained with the same simplification as 
Eq. (\ref{eq:res}) and for $m_t\ll m$. Numerically, 
for relistic values of $t$ and soft-breaking parameters, 
this result does not differ very much from 
Eq. (\ref{eq:nedm}) and finally we 
conclude that the EDM of the neutron in 
constrained MSSM with three generations 
does not exceed the nonsupersymmetric SM predictions. 

\section{MSSM with four generations and neutron EDM}

The extreme smallness of the result (\ref{eq:nedm}) 
kills any hopes of getting any 
observable effect in the model. Moreover, it 
never exceeds the 
non-supersymmetric SM result. The situation is 
somehow different in the 
constrained MSSM with four generations of fermions. 
This model acquired 
significant attention in recent years, since its 
parameter space is already 
severely restricted by the existing data \cite{GMP,CHW}. 
The detailed analysis 
of the renormalization group equations in Ref. \cite{GMP} 
showed that the condition 
of perturbative evolution for all Yukawas requires 
the masses of the fourth 
generation quarks to be under $200$ GeV. The 
similarity between $m_t$ and 
$m_{t'}$ suggests that the third and fourth 
generations are strongly mixed, 
possibly with the angle $V_{tb'}\sim {\cal O}(\lambda)$ 
or even $\sim {\cal O}(1)$. This means that possible 
CP-odd invariant combination of angles 
involving second, third and fourth generations, 
$\Ima(V^*_{ts}V^{\,}_{tb'}V^*_{cb'}V^{\,}_{cs})$ is bigger 
than $J_{CP}$ in SM by a factor 
of the order $\lambda^{-1}-\lambda^{-2}$ (See also Ref. \cite{HP1}). 

Clearly, this is not the main source of enhancement 
for CP-odd observables in the model. A tremendous 
enhancement is associated now with the change of 
masses and Yukawa couplings. Instead of being 
proportional to $m_c^2m_t^4Y_b^2$, 
all the results for EDMs contain now the factor 
$m_t^2m_{t'}^2(m_t^2-m_{t'}^2)Y_{b'}^2$. 
Obviously, this factor is comparable 
with $m_{scalar}^6$ and independent from 
all "light" flavours: $u$, $d$, $s$, $c$ and $b$.  

The detailed calculation of EDMs in the constrained 
MSSM with four generations 
is more difficult than in the conventional 
three-generation model. The main 
complication now is that not only $m_t$, but also 
$m_{t'}$ and $m_{b'}$ are 
comparable with $m$, so that we have to hold 
the exact dependence of these masses. The language 
of mass insertions is completely inadequate now. Instead, 
we have to perform the analysis in the mass 
eigenstate approach. In the limit of 
small $t$ it is still possible to obtain the result 
in a closed analytical form. 
Bearing in mind, however, the big degree of uncertainty 
related with unknown masses and mixing angles, 
we believe that simple estimates are sufficient in 
this case. In particular, we can say that the 
renormalization group evolution of the Yukawa couplings and 
soft breaking parameters from the unification scale 
introduces at the electroweak scale CP-violating 
phases of the order $\Ima(V^*_{ts}V^{\,}_{tb'}V^*_{cb'}
V^{\,}_{cs})Y_t^2Y_{t'}^2(Y_t^2-Y_{t'}^2)Y_{b'}^2$. 
Thus, we can use the estimates of the EDM in the 
presence of CP-violating phases 
with the simple substitution: 
$\sin(\phi_{A(B)})\rightarrow
\Ima(V^*_{ts}V^{\,}_{tb'}V^*_{cb'}V^{\,}_{cs})$. 
Then for the chromoelectric dipole moment 
of the s-quark we have:
\be
\tilde d_s\sim 
g_s\fr{5\alpha_s}{72\pi}~\fr{m_s}{m^2}~\Ima(V^*_{ts}
V^{\,}_{tb'}V^*_{cb'}V^{\,}_{cs})
\ee
The resulting estimate for the electric dipole 
moment of the neutron now is:
\be
d_N\sim \Ima(V^*_{ts}V^{\,}_{tb'}V^*_{cb'}V^{\,}_{cs})
\left(\fr{200\mbox{GeV}}{m}\right)^2 2.5\cdot 10^{-23} e\cdot cm
\label{eq:4n}
\ee
where we took the supersymmetric mass to be of 
the order of $200$ GeV. The 
overall enhancement factor in the MSSM with 
four generations in comparison with 
the three-generation case is enormous, being 
roughly of the order $10^{8}$. As a 
result, the EDM of the neutron is predicted 
in this model to be at the 
measurable level since we expect 
$\Ima(V^*_{ts}V^{\,}_{tb'}V^*_{cb'}V^{\,}_{cs})\sim 
\lambda^5-\lambda^4\sim 10^{-4}-10^{-3}$. 
The current experimental limit on the EDM of the
neutron translates Eq. (\ref{eq:4n}) into the following 
constraint on the CP-odd combination of matrix elements:
\be
\Ima(V^*_{ts}V^{\,}_{tb'}V^*_{cb'}V^{\,}_{cs})
\left(\fr{200\mbox{GeV}}{m}\right)^2\leq 4\cdot 10^{-3}
\ee

It is instructive to compare MSSM with four 
generations and nonsupersymmetric SM 
with the same number of generations. It turns 
out that in SM with four 
generations the result arises at three-loop 
level and these loops are not 
supported by the large logarithmic factors. 
It means that the phase space factor 
in the denominator is large, confining neutron 
EDM to lie within $10^{-29}e\cdot 
cm$ \cite{HP,HP1}. From the point of view of 
the parameters defined at the 
electroweak scale, the EDM in the MSSM with 
four generations arises just at 
one-loop level and therefore is considerably 
larger than in the non-supersymmetric case.

\section{\bf Conclusions}
\label{concl}

We have shown that the limit on the colour 
EDM of the strange quark obtained in 
\cite{KKZ} implies that the constraints
on the phases $\phi_A$ and $\phi_B$ in the soft-breaking sector of MSSM
are stronger than those from the 
quark EDM channel \cite{MSSM-EDM}.
This strengthens the so called supersymmetric CP-problem 
and may have an impact on some 
applications of the specific SUSY CP-violation 
which use the phases on the edge 
of the previous limits \cite{SUSY-CP1}. The 
solution to the supersymmetric 
CP-problem may be related, see 
for example Ref. \cite{MR}, with the dynamical 
suppression of gaugino masses and/or $A$ parameter. 
Alternatively, all phases can be chosen zero 
at the unification scale \cite{DGH}. 

In the MSSM with the CP-violation coming from 
the Yukawa matrices simple order-of-magnitude 
estimates and the numerical calculations are known to 
disagree by many orders of magnitude. We 
demonstrate that analytical results 
are possible for EDMs in the limit of 
small $t$. The chargino-exchange 
diagram yields non-vanishing EDMs for 
the down-quark sector in the lowest 
possible (first) order in $t$. The result 
shows explicitly the suppression factor 
associated with Yukawa couplings and 
basically confirms the order-of-magnitude 
estimates \cite{DGH,RS}. It provides the 
possibility to check all numerical 
calculations by lowering the high-energy 
scale. We believe, however, that even 
for  $t\simeq 1/2$, corresponding to 
$\Lambda\sim  10^{19}$ GeV the results 
cannot differ from ours by many orders 
of magnitude. The contrary would be 
possible only if a kind of singularity exists 
in the solutions of the renormalization group equations 
at some point between $M_W$ and $\Lambda$. We 
reject this possibility and therefore conclude 
that the EDM of the neutron in the 
constrained MSSM does not exceed its 
SM prediction~\cite{KhZ}. 

The four-generation modification of the 
MSSM can be viewed as an interesting 
generalization of the conventional MSSM 
and at the same time it can be checked 
and/or ruled out experimentally in the nearest 
future \cite{GMP,CHW}. We 
believe that the similarity between $m_t$ and 
$m_{t'}$ suggests also non-zero, 
presumably large, mixing between the third and 
the fourth generations. At the same time, 
the relevant combination of masses is much 
bigger in this case and the 
enhancement factor is given by
$(m_{b'}^2m_t^2)/(m_{b}^2m_c^2)\sim 10^{8}$. This 
leads to a large EDM of the neutron 
at the measurable level of $10^{-26} e\cdot cm$. 

\section{\bf Acknowledgements}

We would like to thank I.B. Khriplovich 
and A.R. Zhitnitsky for helpful 
discussions. This research was partially funded by 
N.S.E.R.C of Canada.
The work of M.P. is supported by NATO 
Science Fellowship, N.S.E.R.C., grant \#  
189 630 and Russian Foundation for Basic 
Research, grant \# 95-02-04436-a.

\newpage
{\Large\bf Appendix}

{\bf A)}\hskip.3cm Assuming 
that the renormalization group coefficient $t$
is small, the renormalization group equations can be solved 
analyticaly by iterations. In this case, the coefficients
which appear in equation (\ref{eq:I1}) are found to 
have the following explicit form:
\begin{eqnarray}
c_1&=&-t\,
\bigl[m_{Q_0}^2+m_{u_0}^2+m_{H_{u_0}}^2 +A^2\bigl]
+{\cal O}(t^2)+\ldots\nonumber\\
\noalign{\vskip3truept}
f_1&=&-t\,
\bigl[m_{Q_0}^2+m_{d_0}^2+m_{H_{d_0}}^2 +A^2\bigl]
+{\cal O}(t^2)+\ldots\nonumber\\
\noalign{\vskip3truept}
h_1&=&-t+{\cal O}(t^2)+\ldots\\
\noalign{\vskip3truept}
k_1&=&-3t+{\cal O}(t^2)+\ldots\nonumber\\
\noalign{\vskip3truept}
l_1&=&-2t\bigl[m_{Q_0}^2+m_{u_0}^2+m_{H_{u_0}}^2 +
A^2\bigl]+
{\cal O}(t^2)+\ldots\nonumber
\end{eqnarray}
These quantities arise after the first iteration 
and are the relevant entries to the u-squark mass
matrix for the calculation of chargino-exchange
contribution to quark EDM.  

Since gluino-induced quark EDMs appear only in
the third power in the renormalization group coefficient $t$,
we have to perform also the second iteration. 
The results for $c_2$, $f_2$, $c_2'$, $l_2$ 
and $l_2'$ are:
\begin{eqnarray}
c_2&=&3\,t^2\,A^2+{\cal O}(t^3)+\ldots\nonumber\\
\noalign{\vskip3truept}
f_2&=&3\,t^2\,A^2+{\cal O}(t^3)+\ldots\nonumber\\
\noalign{\vskip3truept}
c_2'&=&\,t^2\,A^2+{\cal O}(t^3)+\ldots\\
\noalign{\vskip3truept}
l_2&=&6\,t^2\,A^2+{\cal O}(t^3)+\ldots\nonumber\\
\noalign{\vskip3truept}
l_2'&=&t^2\,\bigl[m_{Q_0}^2+m_{d_0}^2+
m_{H_{d_0}}^2 +3A^2\bigl]+
{\cal O}(t^3)+\ldots\nonumber
\end{eqnarray}
These expressions are obtained starting with 
diagonal soft-breaking mass parameters at the 
unification scale; the index $0$ refers to 
the parameters at that scale. In order to obtain
the coefficients for the entries in the 
d-squark mass matrix, one has only to interchange
the indices $u$ and $d$.

{\bf B)}\hskip.3cm We would also like to 
present the complete solutions up to the 
second order in $t$ for the renormalization group
equations associated with the soft-breaking
parameters. We used the notations and the 
one-loop $\beta$-functions given in \cite{SUSY-CP}. 
\begin{eqnarray}
{\bf m}^2_{Q_2}=\hskip-5mm& &m^2_{Q_2}+
a^q{\bf Y}^{\dagger}_u{\bf Y}_u
+b^q{\bf Y}^{\dagger}_d{\bf Y}_d+\nonumber\\
& &+t^2\,A^2 \big [ 3 ({\bf Y}^\dagger_{u_2}
{\bf  Y}_{u_2})^2 + 
3({\bf Y}^\dagger_{d_2} {\bf Y}_{d_2})^2 + 
\{ {\bf Y}^\dagger_{u_2} {\bf Y}_{u_2},\,
{\bf Y}^\dagger_{d_2} {\bf Y}_{d_2}\}\big ]\nonumber
\end{eqnarray}
\eject
\begin{eqnarray}
{\bf m}^2_{u_2}=\hskip-5mm& &m^2_{u_2}+
a^u{\bf Y}_u {\bf Y}^{\dagger}_u+\nonumber\\
& &+\frac{1}{2}t^2 \bigl[ 12 A_0^2 
{\bf Y}_{u_2}{\bf Y}^\dagger_{u_2}
{\bf Y}_{u_2}{\bf Y}^\dagger_{u_2}+ 
(4 A_0^2 + b^u ) {\bf Y}_{u_2}{\bf Y}^\dagger_{d_2}
{\bf Y}_{d_2}{\bf Y}^\dagger_{u_2}\bigl]\nonumber\\
\noalign{\vskip4truept}
{\bf m}^2_{d_2}=\hskip-5mm& &m^2_{d_2}+
a^d{\bf Y}_d {\bf Y}^{\dagger}_d+\nonumber\\
& &+\frac{1}{2}t^2 \bigl[ 12 A_0^2 
{\bf Y}_{d_2}{\bf Y}^\dagger_{d_2}
{\bf Y}_{d_2}{\bf Y}^\dagger_{d_2} + 
(4 A_0^2 + b^d ) {\bf Y}_{d_2}{\bf Y}^\dagger_{u_2} 
{\bf Y}_{u_2}{\bf Y}^\dagger_{d_2}\bigl]\\
\noalign{\vskip4truept}
{\bf h}_{u_2}=\hskip-5mm& &{\bf Y}_{u_2}\Big[A_u
+a^{h_u}{\bf Y}^{\dagger}_u{\bf Y}_u+
b^{h_u}{\bf Y}^{\dagger}_d{\bf Y}_d\Big]\nonumber\\
\noalign{\vskip4truept}
{\bf h}_{d_2}=\hskip-5mm& &{\bf Y}_{d_2}\Big[A_d
+a^{h_d}{\bf Y}^{\dagger}_d{\bf Y}_d+
b^{h_d}{\bf Y}^{\dagger}_u{\bf Y}_u\Big]\nonumber
\end{eqnarray}

In the above formulae we used the following notations:
\begin{eqnarray}
a^q=\hskip-5mm& &-t\,\bigl[m_{Q_0}^2+
m_{u_0}^2+m_{H_{u_0}}^2 +A^2\bigl]+\frac{1}{4}\Big[
\int\limits_0^{-t} (m_{{H_u}_1}^ 2 - 
m_{{H_u}_0}^2){\sl d} t+\nonumber\\
& & + t^2 [\rho_{Q_0}^2 + \rho_{u_0}^2 + 
A\eta_{u_0}] - t^2 \bigl[2m_{Q_0}^2+
2m_{u_0}^2+2m_{H_{u_0}}^2  + 3A_0^2\bigl]\Big]\nonumber\\
\noalign{\vskip4truept}
b^q=\hskip-5mm& &-t\,\bigl[m_{Q_0}^2+
m_{d_0}^2+m_{H_{d_0}}^2 +A^2\bigl]+
\frac{1}{4}\Big[\int\limits_0^{-t} (m_{{H_d}_1}^ 2 - 
m_{{H_d}_0}^2){\sl d} t+\nonumber\\
& &+ t^2 [\rho_{Q_0}^2 + \rho_{d_0}^2 + 
A\eta_{d_0}] - t^2 \bigl[2m_{Q_0}^2+
2m_{d_0}^2+2m_{H_{d_0}}^2 +3A^2\bigl]\Big]\nonumber \\
\noalign{\vskip4truept}
a^{u,d}=\hskip-5mm& &-2t\,[m_{Q_0}^2+
m_{{u,d}_0}^2+m_{H_{{u,d}_0}}^2 +A^2\bigl]+
\int\limits_0^t (m_{{H_{u,d}}_1}^ 2 - 
m_{{H_{u,d}}_0}^2){\sl d} t+\\
& &+ \frac{1}{2}t^2\Big[\rho_{Q_0}^2 +A\eta_{{u,d}_0}
-2\beta_{{u,d}_0}\bigl[m_{Q_0}^2+
m_{{u,d}_0}^2+m_{H_{{u,d}_0}}^2 +A^2\bigl]\Big]\nonumber\\
\noalign{\vskip4truept}
b^{u,d}=\hskip-5mm& &2\,\bigl[m_{Q_0}^2+m_{{d,u}_0}^2+
m_{H_{{d,u}_0}}^2 +A^2\bigl]\nonumber\\
\noalign{\vskip4truept}
a^{h_{u,d}}=\hskip-5mm& &-3t\,A+\frac{1}{4}
t^2\,\bigl[-8A\beta_{{u,d}_0}+3\eta_{{u,d}_0}\bigl]
\nonumber\\
\noalign{\vskip4truept}
b^{h_{u,d}}=\hskip-5mm& &-tA+\frac{1}{4}t^2\,
\bigl[A(\beta_{{u,d}_0}
-4\beta_{{d,u}_0})+{1\over 2}(2\eta_{{d,u}_0}-
\eta_{{u,d}_0})\bigl]\nonumber
\end{eqnarray}
Finally, $A_{u,d}$, $m_{Q_2}$, $m_{u_2}$ and
$m_{d_2}$ are soft-breaking parameters renormalized
by gauge interactions and flavour-conserving 
Yukawa interactions. The indices $0$, $1$ and $2$
refer to the iteration to be considered
for the corresponding quantity. The functions
$\eta$, $\rho$ and $\beta$ are scalar quantities involving
gauge couplings, gaugino masses and 
Tr$\bigl[{\bf Y}_i^\dagger{\bf Y}_i\bigl]$.
Their exact form can be extracted from Ref.\cite{SUSY-CP}.
Using these solutions, one can identify all the
coefficients in equation (\ref{eq:I1}) up
to the second power in $t$.

\newpage
{\bf Figure captions}
\bigskip

\noindent
{\bf Fig. 1.} Quark self-energy involving chargino exchange, generating EDM (CEDM) of quark in the external electromagnetic (color) field. 

\medskip
\noindent
{\bf Fig. 2.} Quark self-energy involving gluino exchange, generating EDM (CEDM) of quark in the external electromagnetic (color) field. 

\bigskip\bigskip\bigskip
\bigskip\bigskip\bigskip

\begin{figure}[hbtp]
\begin{center}
\mbox{\epsfxsize=9truecm
\epsffile{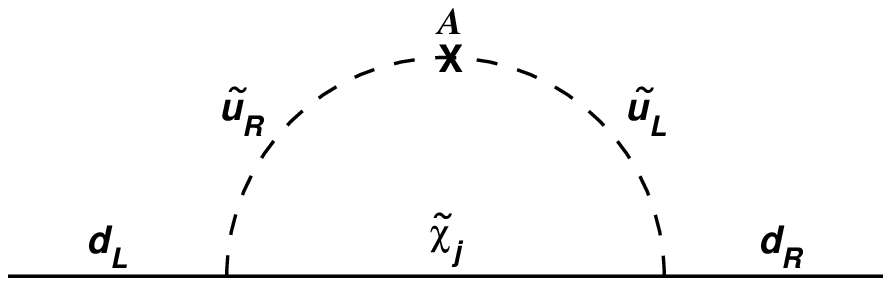}}
\end{center}
\end{figure}
\centerline{{\bf Fig. 1.}}

\bigskip\bigskip\bigskip
\bigskip\bigskip\bigskip

\begin{figure}[hbtp]
\begin{center}
\mbox{\epsfxsize=9truecm
\epsffile{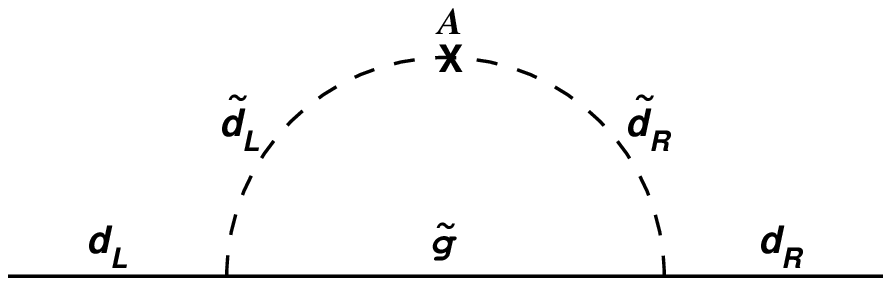}}
\end{center}
\end{figure}
\centerline{{\bf Fig. 2.}}


\begin{thebibliography}{99}
\bi{LG} K.F. Smith et al, Phys. Lett. {\bf B234} (1990) 191;
\\I.S. Altarev et al, Phys. Lett. {\bf B276} (1992) 242.

\bibitem{habnil}
H.E. Haber and G.L. Kane,  Phys. Rep. {\bf 117} (1985) 75;\\
H.P. Nilles,  Phys.Rep. {\bf 110} (1984) 1.


\bi{MSSM-EDM} J. Ellis, S. Ferrara and 
D.V. Nanopoulos, Phys. Lett. {\bf B114} 
(1982) 231;\\
W. Buchmuller and D. Wyler, Phys. Lett. {\bf B121} (1983) 321;\\
J. Polchinski and M.B. Wise, Phys. Lett. {\bf B125} (1983)
393;\\
F. del Aquila {\em et al.}, Phys. Lett. {\bf B126} (1983) 71.

\bi{DGH} M. Dugan, B. Grinstein and L. Hall, Nucl. Phys. {\bf B255} (1985) 413.

\bi{Dunc} M.J. Duncan, Nucl. Phys. {\bf B221} (1983) 285.

\bi{KhZ}M.B. Gavela {\em et al.}, 
Phys. Lett. {\bf B109} (1982) 215; \\
I.B. Khriplovich and A.R. Zhitnitsky, 
Phys. Lett. {\bf B109} (1982) 490; \\
X.-G. He, B.J.H. McKellar and S. Pakvasa, 
Int.J.Mod.Phys. {\bf A4} (1989) 5011;\\
ERRATUM-ibid.{\bf A6} (1991) 1063.

\bi{RS} A. Romanino and A. Strumia, hep-ph/9610485.

\bi{Ital} S. Bertolini and F. Vissani, 
Phys. Lett. {\bf B324} (1994) 164.

\bi{Japan} T. Inui {\em et al.}, 
Nucl. Phys. {\bf B449} (1995) 49.

\bi{Engl} S.A. Abel, W.N. Cottingham and I.B. Wittingham, 
Phys. Lett. {\bf B370} (1996) 106. 

\bi{HP} C. Hamzaoui and M.E. Pospelov, 
Phys. Lett. {\bf B357} (1995) 616.

\bi{HP1} C. Hamzaoui and M.E. Pospelov, 
Phys.Rev. {\bf D54} (1996) 2194.

\bi{GMP} J.F. Gunion, D.W. McKay and H. Pois, 
Phys.Rev. {\bf D53} (1996) 1616;\\ 
Phys. Lett. {\bf B334} (1994) 339.

\bi{CHW} M. Carena, H.E. Haber and C.E.M. Wagner 
Nucl.Phys. {\bf B472} (1996) 55.

\bi{EF} J. Ellis and R. Flores, Phys. Lett. {\bf B377} (1996) 83.

\bi{JJ} R.L. Jaffe and X.-D. Ji, 
Phys. Rev. Lett. {\bf 67} (1991) 552.

\bi{Pol} H.-C. Kim, M.V. Polyakov and K.Goeke, 
Phys. Lett. {\bf B387} (1996) 577.

\bi{Hats} S. Aoki, M. Doui, T. Hatsuda and Y. Kuramashi, 
hep-lat/9608115.

\bi{Ji} H.-X. He and X.-D. Ji, Phys.Rev. {\bf D54} (1996) 6897.

\bi{KKZ} V.M. Khatsimovsky, I.B. Khriplovich and A.R.
Zhitnitsky, Z. Phys. {\bf C36} (1987) 455.

\bi{KKY} V.M. Khatsimovsky, I.B. Khriplovich and 
A.S. Yelkhovsky, Ann. Phys. {\bf 186} (1988) 1.

\bi{KK} V.M. Khatsimovsky and I.B. Khriplovich, 
Phys. Lett. {\bf B296} (1994) 219.

\bi{KZy} I.B. Khriplovich and K.N. Zyablyuk, 
Phys. Lett. {\bf B383} (1996) 429.

\bi{Z} A.R. Zhitnitsky, hep-ph/9604314, to appear in Phys.Rev. {\bf D}

\bi{Jarlsk} C. Jarlskog, Phys.Rev.Lett. {\bf 55} (1985) 1039.

\bi{Shab} E.P. Shabalin, Yad.Fiz. {\bf 28} (1978) 151 
(Sov. J. Nucl. Phys. {\bf 28} (1978) 75).

\bi{SUSY-CP1} S.A. Abel and J.-M. Fr\`ere, Phys. Rev. 
{\bf D55} (1997) 1623.

\bi{MR} R.N. Mohapatra and A. Riotto, Phys.Rev. {\bf D55}
(1997) 1138.

\bi{SUSY-CP} S.P. Martin and M.T. Vaughn, Phys.Rev. {\bf D50}
(1994) 2282.




\end{thebibliography}
\end{document}